\documentclass{appolb}
\usepackage{graphicx}
\usepackage{hyperref}

\begin{document}
\newcommand{\ksta}{$K^{*0}$}
\newcommand{\kst}{$K^{*0}$~}
\newcommand{\akst}{$\overline{K}^{*0}$~}
\newcommand{\aksta}{$\overline{K}^{*0}$}
\newcommand{\kzs}{\mbox{$K^0_S$}~}

\newcommand{\psia}{\psi_{\rm A}}
\newcommand{\psic}{\psi_{\rm C}}
\newcommand{\PsiRP}{\ensuremath{\Psi_{\rm RP}}\xspace}
\newcommand{\psisp}{\Psi_{\rm SP}}
\newcommand{\Rep}{\ensuremath{R_{\rm SP}^{(1)}}\xspace}

\def \psitwo {\Psi_{\rm 2}}
\def \pzstwo {P_{\rm z,\,{\rm s2}}}

\newcommand{\ph}{\ensuremath{P_{\rm H}}\xspace}
\newcommand{\pt}{p_{\rm T}}
\newcommand{\pT}{\ensuremath{p_{\rm T}}\xspace}

\newcommand{\sNN}{\ensuremath{\sqrt{s_{_{\rm NN}}}}\xspace}
\newcommand{\mt}{\mbox{$m_T$}\xspace}
\newcommand{\Et}{\mbox{${\rm E}_T$}\xspace}
\newcommand{\sqsn}{\mbox{$\sqrt{s_{_{NN}}}$}\xspace}
\newcommand{\apt}{\mbox{$\langle p_T \rangle$}\xspace}
\newcommand{\costhe}{\mbox{$\langle\cos\theta_p^{\ast}\rangle$}\xspace}

\def \psitwoep {\Psi_{\rm 2}^{\rm EP}}

\newcommand{\Qv}{{\rm \bf Q}}
\newcommand{\tpspec}{{\rm t,p}}
\newcommand{\Xpm}{\Qx^\tpspec}
\newcommand{\Ypm}{\Qy^\tpspec}
\newcommand{\Qtp}{\Qv^\tpspec}
\newcommand{\PsiSPtp}{\psisp^\tpspec}

\newcommand{\Qx}{Q_{\rm x}}
\newcommand{\Qy}{Q_{\rm y}}

\newcommand{ \la }{\langle}
\newcommand{ \ra }{\rangle}
\newcommand{ \bs }{{\bf s}}
\newcommand{ \bl }{{\bf l}}
\newcommand{ \bv }{{\bf v}}
\newcommand{ \bP }{{\bf P}}
\newcommand{ \bp }{{\bf p}}
\newcommand{ \bq }{{\bf q}}
\newcommand{ \bz }{{\bf z}}
\newcommand{ \bB }{{\bf B}}
\newcommand{ \bJ }{{\bf J}}

\newcommand{ \rT }{{\rm T}}

\newcommand{ \pd }{{\partial}}
\newcommand{ \eps }{\ensuremath{\varepsilon}\xspace}
\newcommand{ \ds }{\displaystyle}
\newcommand{\mean}[1]{\la #1 \ra}
\newcommand{ \Ahat} {\hat{A}}
\newcommand{ \Eps} {\mathcal{E}}
\newcommand{ \grad }{{\bf \nabla}}
\newcommand{ \curl }{{\rm \bf curl\, }}
\newcommand{ \bomega }{{\boldsymbol{\omega}}}
\newcommand{ \lam }{{$\Lambda$}\xspace}
\newcommand{ \alam }{{$\bar{\Lambda}$}\xspace}

\newcommand{\PsiEP}{\Psi_{\rm 1,EP}}
\newcommand{\PsiEPn}{\Psi_{\rm n,EP}}
\newcommand{\REP}{R_{\rm 1,EP}}
\newcommand{\REPt}{R_{\rm 2,EP}}
\newcommand{\REPn}{R_{\rm n,EP}}

\newcommand{\vtp}{v_1\{\PsiSPtp\}}
\newcommand{\vp}{v_1\{\PsiSPp\}}
\newcommand{\vt}{v_1\{\PsiSPt\}}

\newcommand{\Minv}{\ensuremath{\mathrm{M}_\mathrm{inv}}\xspace}
\newcommand{\fBG}{\ensuremath{f_\mathrm{BG}(\Minv)}\xspace}

\newcommand{\orange}[1]{\textcolor{orange}{#1}}
\newcommand{\blue}[1]{{\color{blue}{#1}}}
\newcommand{\green}[1]{{\color{dgreen}{#1}}}
\newcommand{\red}[1]{{\color{red}{#1}}}
\newcommand{\magenta}[1]{{\color{magenta}{[#1]}}}

\def \new {\blue}
\def \old {\orange}
\def \ask {\magenta}
\def \mod {\green}
\def \moved {\red}
\newcommand{\note}[1]{\textbf{\textcolor{magenta}{[#1]}}}
\def \new {\blue}

\title{Spin polarization measurements in relativistic heavy-ion collisions%
\thanks{Presented at Quark Matter 2022 (Krakow, Poland)\href{https://indico.cern.ch/event/895086/contributions/4615144/}{link to presentation}}%
}
\author{Debojit Sarkar
\thanks{D.S is supported by the U.S. Department of Energy, Office of Science, Office of Nuclear Physics, Grants No. DE-FG02- 92ER40713 }
\address{Wayne State University (Detroit, USA); \\
Niels Bohr Institute (University of Copenhagen, Denmark)}
}
\maketitle
\begin{abstract}
The hot and dense matter formed in relativistic heavy-ion collisions at Relativistic Heavy Ion Collider (RHIC) and Large Hadron Collider (LHC) is termed quark-gluon plasma (QGP). The evolution of the medium is characterized by non-trivial velocity and vorticity fields, resulting in the polarization of the produced particles. The spin polarization, being sensitive to the hydrothermal (flow velocity and temperature) gradients, is unique compared to conventional observables that are sensitive to the hydrothermal fields only. Hence, the recent measurements of global and local hyperon spin polarization and vector meson spin alignment by the LHC and STAR collaborations provide a unique opportunity to probe the QGP substructure with finer details. 
 \end{abstract}

\section{Hyperon polarization and vector meson spin alignment measurements}
The system created in relativistic heavy-ion collisions at the RHIC and LHC might retain a significant fraction of the initial orbital angular momentum of the colliding system through the initial shear in the longitudinal velocity profile of the participants~\cite{Becattini_Lisa_rev}. This shear can generate vorticity perpendicular to the reaction plane defined by the impact parameter vector and the beam direction. Due to the spin-orbit coupling, the particles emerging from the system with non-zero spin get polarized along the vorticity of the system. This phenomenon is termed global polarization as all particles get polarized along one preferential direction defined by the orbital angular momentum of the colliding system~\cite{Becattini_Lisa_rev, Becattini_voloshin_vort, voloshin_pol}. The global polarization due to vorticity is expected to be the same for both particle and antiparticle in a system with small or zero baryon chemical potential. However, the initial ultra-strong magnetic fields, aligned with the direction of the angular momentum, can polarize particles and anti-particles in an opposite direction due to the opposite sign of their magnetic moments. Hence, the precise measurements of the particle and anti-particle global polarization in heavy-ion collisions can provide valuable insights into the QGP properties.\\
The spin of a particle cannot be measured directly. The parity-violating weak decaying strange ($\Lambda$) and multi strange ($\Xi$, $\Omega$) hyperons are experimentally favorable as their spin direction can be reconstructed from the decay daughter angular distribution which is an odd function and depends on the direction of the angular momentum. The mother hyperon polarization is estimated from the daughter baryon angular distribution in the mother’s rest frame~\cite{Becattini_Lisa_rev}, given by: 
\begin{equation}
4\pi \frac{{\rm d}N}{{\rm d}\Omega^{*}} 
 = 1 + \alpha_{\rm H} P_{\rm H} \cdot { \bf  \hat{p}}_{\rm B}^{*}                
 = 1 + \alpha_{\rm H}P_{\rm H}\cos\theta_{\rm B}^{*},
\label{Eq1}   
\end{equation}
where $P_{\rm H}$ is the polarization vector, ${ \bf \hat{p}}_{\rm B}^{*}$ is the unit vector along the daughter momentum, and $\theta_{\rm B}^{*}$ is the angle between the daughter momentum and the polarization vector. The asterisks denote that the quantities are in the rest frame of the parent. The hyperon decay parameter ($\alpha_{\rm H}$)~\cite{STAR_gpol_nature} characterizes the anisotropy of the decay daughter angular distribution in the parent rest frame. The $\alpha_{\rm H}$ values get updated in the Particle Data Group (PDG) from time to time~\cite{STAR_gpol_nature, STAR_gpol_200, STAR_gpol_3, ALICE_gpol, ALICE_localpol, STAR_localpol}.\\
In general, the polarization is estimated from the ensemble-averaged projection of the daughter baryon momentum direction along the vorticity source~\cite{Becattini_Lisa_rev}. In heavy-ion collisions, the global polarization vector $P_{\rm H}$ along the initial angular momentum is given by~\cite{STAR_gpol_nature}:
\begin{equation}
P_H = \frac{8}{\pi\alpha_H}\frac{\langle\sin(\Psi_{\rm SP}-\phi_B^\ast)\rangle}{\rm Res(\Psi_{\rm SP})},
\end{equation}
where $\phi_B^\ast$ is the azimuthal angle of the daughter baryon in the parent hyperon rest frame. The spectator plane angle $\Psi_{\rm SP}$ is estimated from the deflection direction of the spectator neutrons~\cite{STAR_gpol_nature, ALICE_gpol}, characterizing the orbital angular momentum direction of the colliding system. The Res($\Psi_{\rm SP}$) is the event plane resolution correction~\cite{evplane_res_voloshin}.\\
 \begin{figure}[htb]
\includegraphics[width=4cm, height=3.8cm]{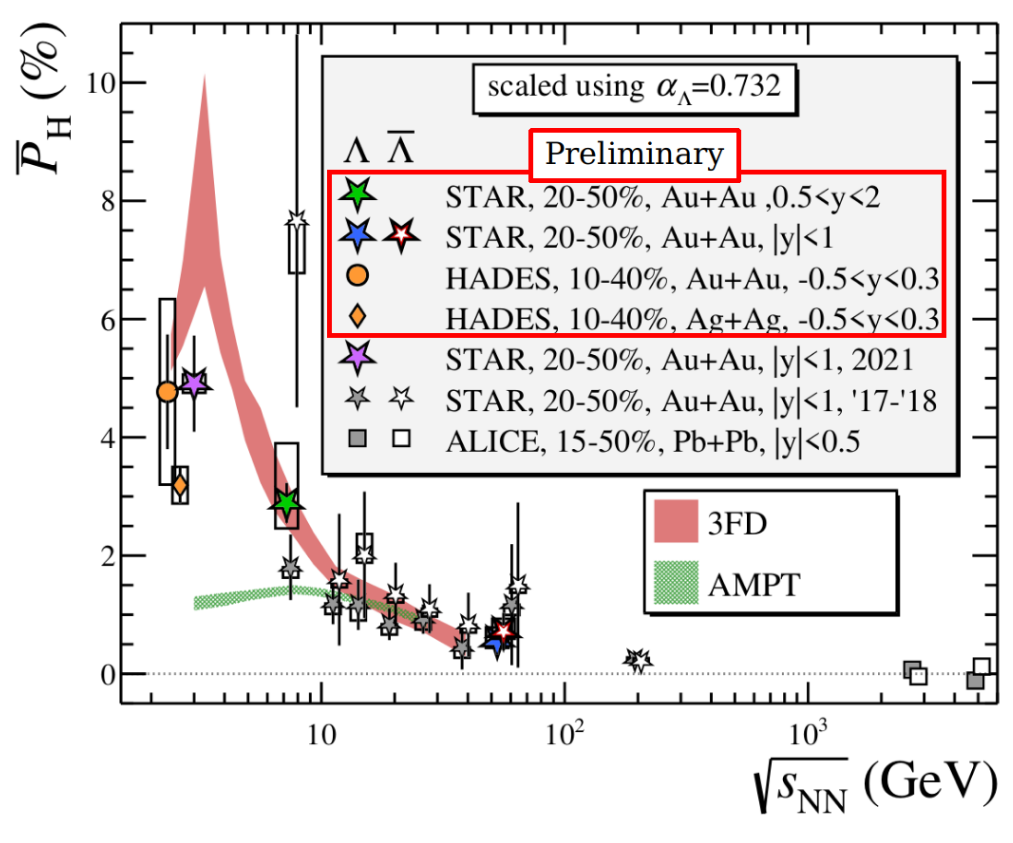}\hspace{0.1cm}
\includegraphics[width=4cm, height=3.7cm]{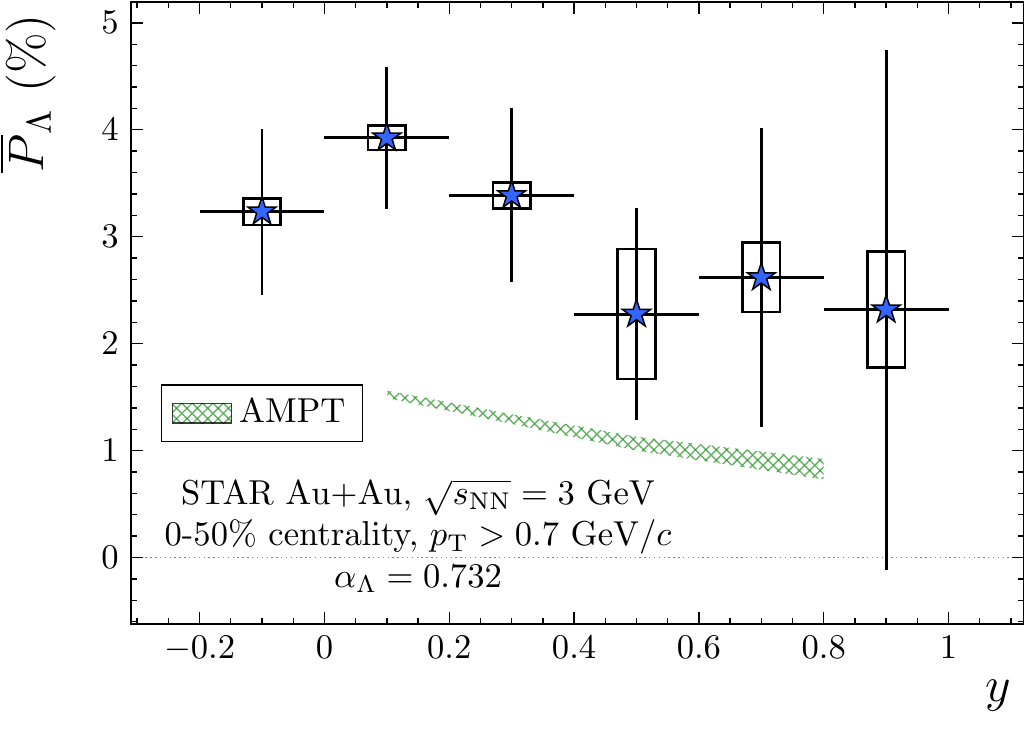}\hspace{0.1cm}
\includegraphics[width=4cm, height=3.9cm]{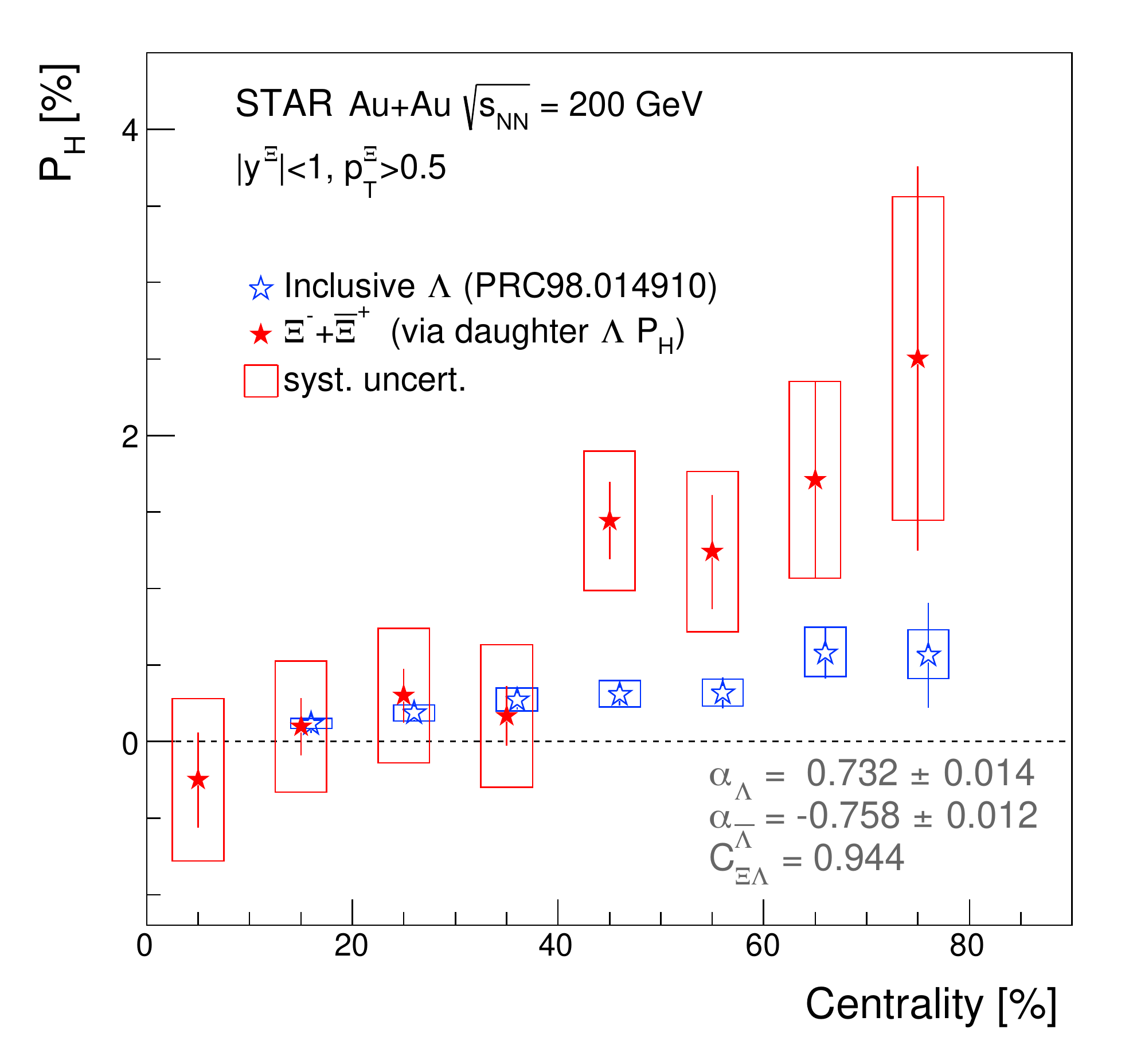}
\caption{ (color online) {\bf Left}:  $\sqrt{s_{_{\rm NN}}}$ dependence of $P_{\Lambda(\overline{\Lambda)}}$ and its comparison with the hydrodynamic (3FD) and transport (AMPT) model estimations~\cite{gpol_fig_joey}. {\bf Middle}: Rapidity dependence of $P_{\Lambda}$ in Au--Au collisions at $\sqrt{s_{_{\rm NN}}} =$~3~GeV with AMPT estimation~\cite{STAR_gpol_3}. {\bf Right}:  Centrality dependence of $P_{\Lambda, \Xi}$ in Au--Au collisions at $\sqrt{s_{_{\rm NN}}} =$~200~GeV~\cite{STAR_Xi_Omega_gpol_200}.}
\label{gpol_lam}
\end{figure}
The STAR collaboration measured significant global polarization of $\Lambda$ ($P_{\Lambda}$) and  $\overline{\Lambda}$ ($P_{\overline{\Lambda}}$) hyperons in the collision energy range $\sqrt{s_{_{\rm NN}}} =$ 3--200~GeV~\cite{STAR_gpol_nature, STAR_gpol_200, STAR_gpol_3}. The polarization magnitude decreases monotonically with $\sqrt{s_{_{\rm NN}}}$ from a few to a fraction of a percent, shown in Fig.~\ref{gpol_lam}. The $P_{\Lambda}$ and $P_{\overline{\Lambda}}$ are consistent with each other within experimental uncertainty. However, $P_{\overline{\Lambda}}$ is systematically higher than $P_{\Lambda}$, and the difference is more prominent at lower $\sqrt{s_{_{\rm NN}}}$. Initial magnetic field and higher baryon chemical potential at lower $\sqrt{s_{_{\rm NN}}}$ can contribute to this difference~\cite{Becattini_Lisa_rev}. More precision measurements and data-model comparisons are required to understand the origin of this behavior. The $P_{\Lambda(\overline{\Lambda})}$ measured by the ALICE Collaboration for Pb-Pb collisions at $\sqrt{s_{_{\rm NN}}} =$ 2.76 and 5.02~TeV~\cite{ALICE_gpol} is consistent with zero within experimental uncertainties (Fig.~\ref{gpol_lam}), following the decreasing trend of $P_{\rm H}$ with increasing $\sqrt{s_{_{\rm NN}}}$. However, hyperon global polarization and slope of the charged particle directed flow ($v_{1}$) at midrapidity, both related to the initial vorticity or tilt in the system, are found to be strongly correlated in the RHIC BES measurements~\cite{ALICE_gpol}. The extrapolation of those measurements, combined with the experimentally observed three times smaller $v_{1}$ in Pb–Pb collisions at $\sqrt{s_{_{\rm NN}}} =$ 2.76~TeV compared to the same in  Au–Au collisions at $\sqrt{s_{_{\rm NN}}} =$ 200~GeV~\cite{ALICE_dirflow}, predict a small but finite $P_{\rm H}$ ($\approx$0.04\%) at the LHC energies~\cite{ALICE_gpol}. The Run-3 data at the LHC with higher statistics will provide a more precise estimate of global vorticity at the LHC energies. \\
Hydrodynamic (3FD)~\cite{gpol_3FD_cal_1} and transport (AMPT)~\cite{gpol_AMPT_cal} calculations reasonably explain the $\sqrt{s_{_{\rm NN}}}$ dependence of $P_{\rm H}$ for $\sqrt{s_{_{\rm NN}}}\geq7.7$~GeV as shown in Fig.~\ref{gpol_lam}. These model calculations estimate the particle polarization from the thermal vorticity~\cite{Becattini_voloshin_vort} at the freeze-out surface assuming local thermodynamic equilibrium of the spin degrees of freedom. It should be noted that the exact nature of the spin-orbit interaction and the spin relaxation times are not known. An extension of the Cooper–Frye formalism is used to calculate the polarization (particle property) from the thermal vorticity (fluid property) generated by hydrodynamic and transport models in all such calculations~\cite{Becattini_Lisa_rev}.
\\ 
Fig.~\ref{gpol_lam} shows a monotonic increase of $P_{\rm H}$ as $\sqrt{s_{_{\rm NN}}}$ decreases. However, $P_{\rm H}$ is expected to vanish at $\sqrt{s_{_{\rm NN}}}=2m_\mathrm{N}$ due to the lack of system angular momentum~\cite{STAR_gpol_3, gpol_upperlimit_3gev}.  Therefore, a peak of $P_{\rm H}$ is expected in the range $2m_{\rm N}\approx1.9<\sqrt{s_{_{\rm NN}}}<7.7$~GeV~\cite{STAR_gpol_3}. The recent model calculations predict this maximum of $P_{\rm H}$ to be around $\sqrt{s_{_{\rm NN}}}\approx3$~GeV~\cite{STAR_gpol_3, gpol_upperlimit_3gev}. The STAR Collaboration measured the $P_{\Lambda}$ at $\sqrt{s_{_{\rm NN}}}=$ 3~GeV, the largest $P_{\Lambda}$ ever measured in the experiment as shown in Fig.~\ref{gpol_lam}~\cite{STAR_gpol_3}. The $P_{\Lambda}$ in Au+Au collisions at $\sqrt{s_{_{\rm NN}}}=$ 2.42~GeV (HADES) is consistent with STAR $\sqrt{s_{_{\rm NN}}}=$ 3~GeV result within measurement uncertainty, preventing a conclusion about the possible maximum of $P_{\Lambda}$ at $\sqrt{s_{_{\rm NN}}}=$ 3~GeV. The $P_{\Lambda}$ in Ag+Ag collisions at $\sqrt{s_{_{\rm NN}}}=$ 2.55~GeV (HADES) is lower than the same measured in Au+Au collisions at $\sqrt{s_{_{\rm NN}}}=$ 2.42~GeV~(HADES) and 3~GeV~(STAR) due to lower value of initial angular momentum in the Ag+Ag system (Fig.~\ref{gpol_lam}). The 3FD hydro calculation~\cite{gpol_3FD_cal_1} seems to work better at lower $\sqrt{s_{_{\rm NN}}}$ range whereas the AMPT~\cite{gpol_AMPT_cal} works well only for $\sqrt{s_{_{\rm NN}}}\geq7.7$~GeV as seen in Fig.~\ref{gpol_lam}. The observation of large $P_{\rm H}$ at $\sqrt{s_{_{\rm NN}}}\leq7.7$~GeV indicate that the hadron gas may support enormous vorticity at low collision energies~\cite{STAR_gpol_3}. On the other hand, the decrease of $P_{\rm H}$ at mid-rapidity with the increase in $\sqrt{s_{_{\rm NN}}}$ can be attributed to the migration of vorticity toward forward rapidity~(y) due to the longitudinal boost invariance at mid-rapidity at higher $\sqrt{s_{_{\rm NN}}}$ ~\cite{gpol_3FD_cal_2}. A forward detector upgrade at RHIC and the LHC is necessary to reconstruct hyperons and measure $P_{\rm H}$ at forward rapidities at higher $\sqrt{s_{_{\rm NN}}}$. In this context, the STAR FXT (Fixed Target) results at $\sqrt{s_{_{\rm NN}}}= 3$~GeV is unique as it covers the range $-0.2\leq y \leq  y_{\rm beam}$, reaching the upper limit of $y_\Lambda$ at this collision energy~\cite{STAR_gpol_3}. As shown in Fig.~\ref{gpol_lam}, no rapidity dependence of $P_{\rm H}$ is observed at $\sqrt{s_{_{\rm NN}}}=$ 3~GeV within the measurement uncertainty, and the AMPT can't explain the data. Further precision measurements of $P_{\rm H}$ up to $y_{\rm beam}$ at higher $\sqrt{s_{_{\rm NN}}}$ are necessary to shed light on this topic as the medium properties at higher $\sqrt{s_{_{\rm NN}}}$ are quite different with the boost invariance at mid-rapidity being a better approximation.\\
%
%
The STAR collaboration also measured the global polarization for multi-strange hyperons such as $\Xi$ (spin 1/2) and $\Omega$ (spin 3/2) in Au+Au collisions at $\sqrt{s_{_{\rm NN}}}=$ 200~GeV~\cite{STAR_Xi_Omega_gpol_200}. This measurement provides insight into the global polarization picture based on thermal vorticity that warrants different particles to be polarized in the same direction with polarization magnitudes depending only on their spin~\cite{Becattini_voloshin_vort}. The  $\Xi$ and $\Omega$ hyperons decay in two steps: $\Xi \rightarrow
\Lambda +\pi^-$ ; $\Omega\rightarrow \Lambda +K$  with subsequent decay of $\Lambda \rightarrow p+
\pi^-$. Both steps in such a cascade decay
are parity violating and can be used for independent polarization  
measurement. However, the decay parameter $\alpha_\Omega$ ($=0.0157\pm0.0021$) being very small, measuring $\Omega$ polarization directly from the daughter $\Lambda$ angular distribution is practically impossible~\cite{STAR_Xi_Omega_gpol_200}. The $\Xi$ and $\Omega$ polarization can be estimated from the daughter polarization ($P_{\Lambda}$, from $\Lambda \rightarrow p+
\pi^-$ decay) using a polarization transfer factor ($C_{\Omega\Lambda}$, $C_{\Xi\Lambda}$), assuming polarization transfer from the parent to the daughter~\cite{STAR_Xi_Omega_gpol_200}. Fig.~\ref{gpol_lam} shows that for semi-central and peripheral collisions, $P_{\Xi} > P_{\Lambda}$, despite both being spin 1/2 particles. The measured $P_{\Xi}$ and $P_{\Omega}$ for the collision centrality 20\%-80\% are found to be slightly larger than the $P_{\Lambda}$ and in reasonable agreement with a multi-phase
transport model (AMPT)~\cite{STAR_Xi_Omega_gpol_200}. These measurements indicate that different spin (for $\Omega$), as well as different freeze-out times or regions, may contribute to larger polarization for $\Xi$ and $\Omega$ compared to that of $\Lambda$~\cite{Xi_Omega_gpol_diff_freezeout}. Further precision measurements for different species will be essential to understand the possible effects of particle mass, spin, and magnetic moments on particle polarization in the context of the fluid vorticity-based global polarization picture.

The vector mesons (e.g \kst, $\phi$) decay through the parity conserving strong decay process. In strong decay, the angular distribution of the decay daughters is an even function and depends only on the strength and not on the direction of the system angular momentum~\cite{BMohanty_spinalign, ALICE_spinalign}. Hence, the angular distribution of the decay daughters with respect to a quantization axis (along system angular momentum, normal to the event plane) in the vector meson’s rest frame provides an estimation of the spin alignment~\cite{BMohanty_spinalign}:
\begin{equation}
\frac{\mathrm{d}N}{\mathrm{d}\cos\theta_{\rm B}^{*}} \propto [ 1 - \rho_{00} + \cos^{2}\theta_{\rm B}^{*}(3\rho_{00} - 1 ) ].
\label{eqn1}
\end{equation}
where $\rho_{00}$ is the only independent diagonal element of 3$\times$3 hermitian spin density matrix with unit trace and is used to quantify the vector meson spin alignment~\cite{BMohanty_spinalign}. $\rho_{00}$ corresponds to the probability of finding a vector meson in spin state 0 out of 3 possible spin states of -1, 0 and 1 ($\rho_{11}$, $\rho_{00}$ and $\rho_{-1-1}$). In absence of spin alignment $\rho_{00} =$ 1/3. Any deviation of $\rho_{00}$ from 1/3 is considered an experimental signature of the vector meson spin alignment~\cite{ALICE_spinalign}.\\
\begin{figure}[htb]
\includegraphics[width=4cm, height=3.7cm]{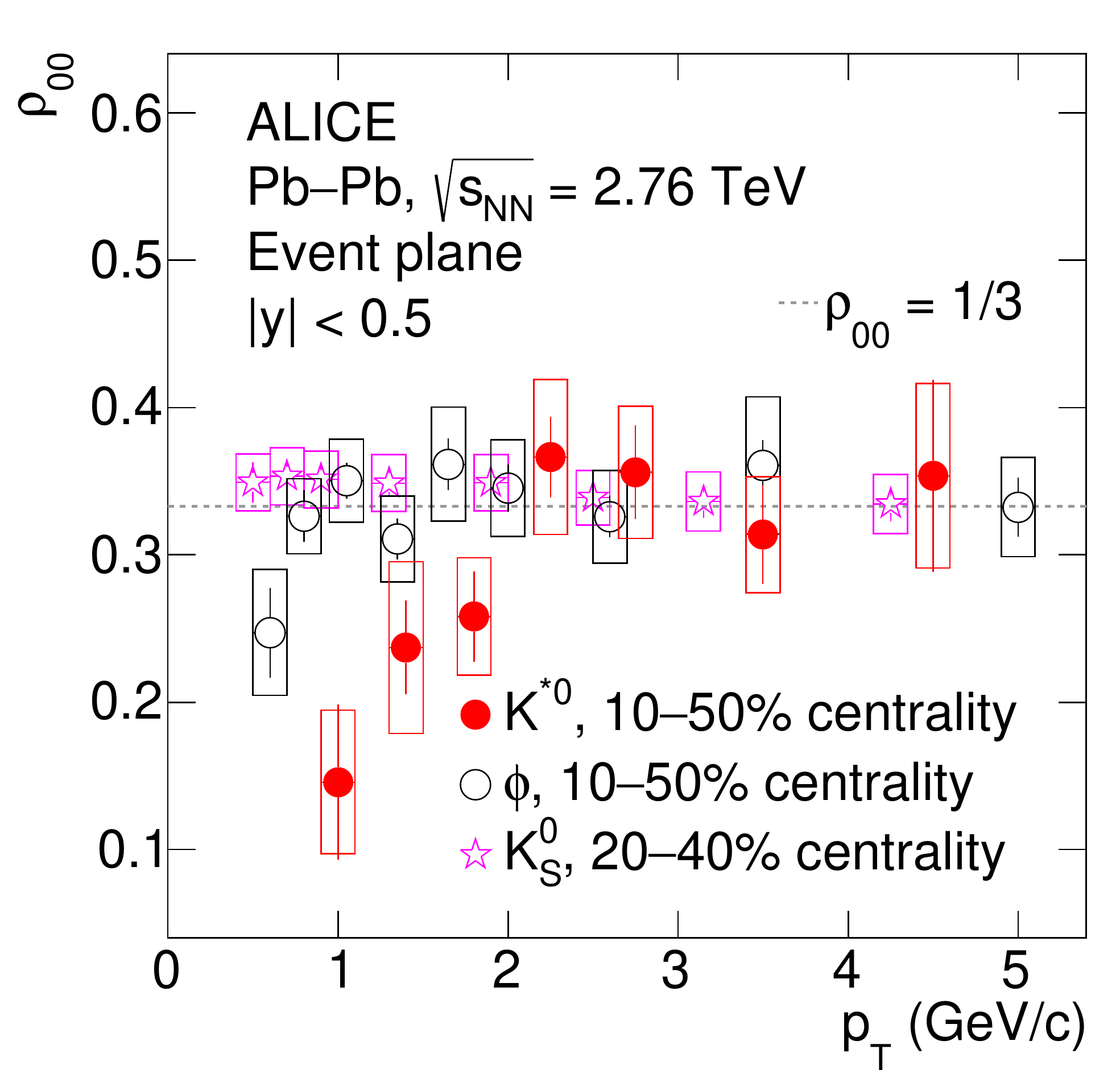}
\includegraphics[width=4cm, height=3.7cm]{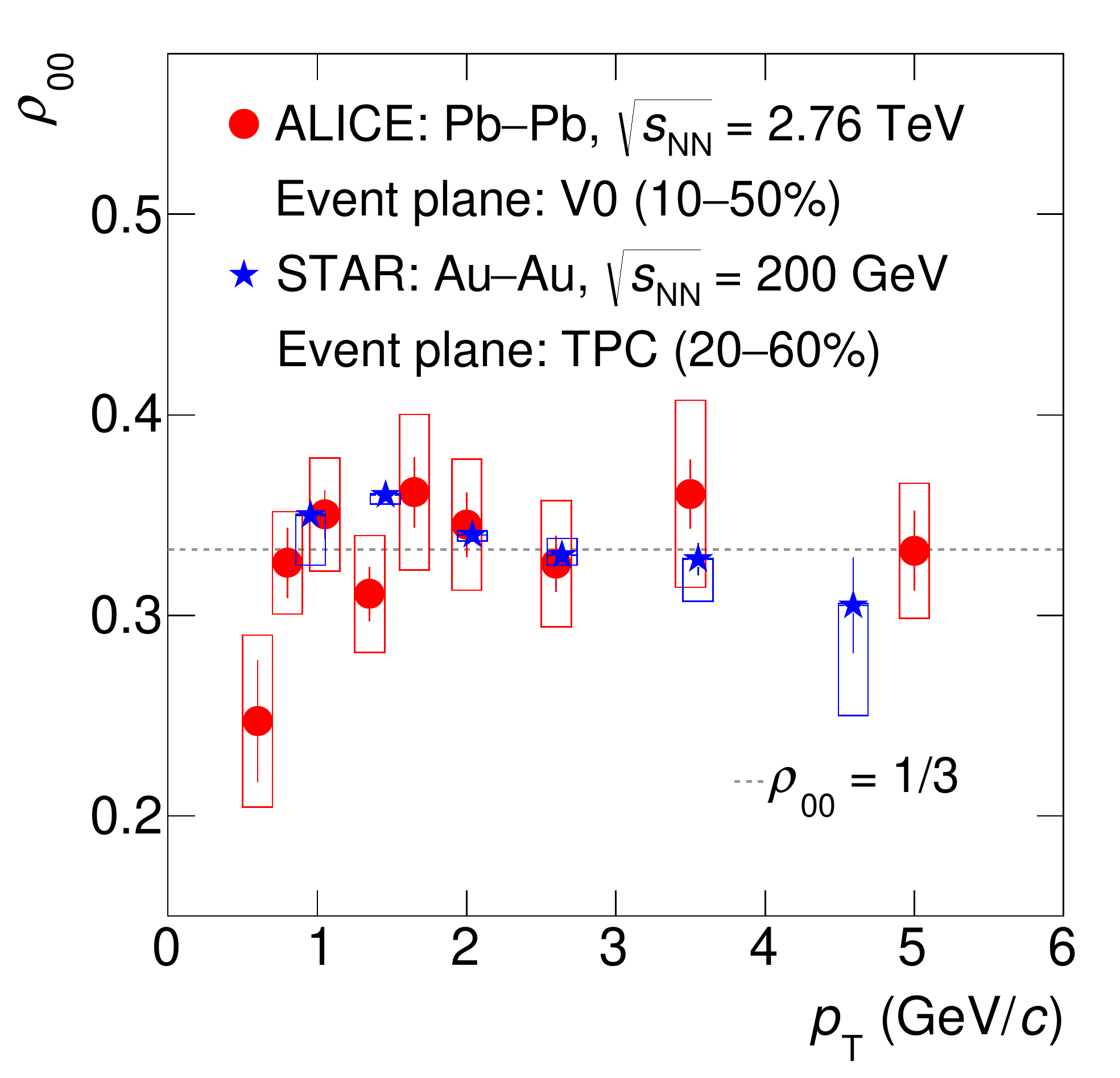}
\includegraphics[width=4cm, height=4.0cm]{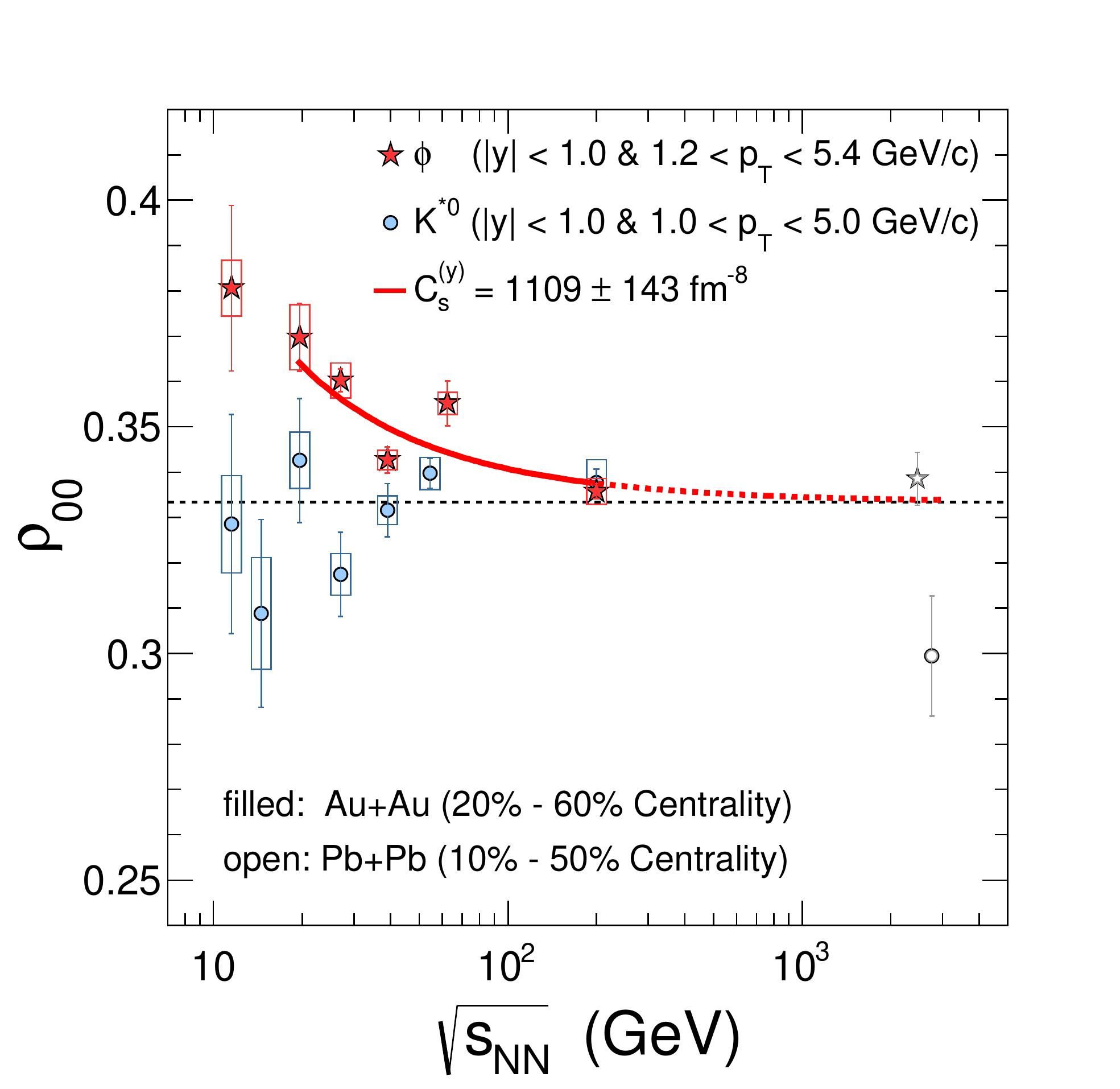}
\caption{ (color online) {\bf Left}: $p_{T}$ dependence of $\rho_{00}$ for \kst and $\phi$ vector mesons and its comparison with \kzs (spin 0) in semi-central Pb-Pb collisions at $\sqrt{s_{_{\rm NN}}} =$ 2.76 TeV~\cite{BMohanty_spinalign}. {\bf {Middle}:}$p_{T}$ dependence of $\rho_{00}$ for $\phi$ in semi-central Pb-Pb (ALICE) and Au-Au collisions (STAR)~\cite{BMohanty_spinalign}. {\bf Right}: $\sqrt{s_{_{\rm NN}}}$ dependence of  $\rho_{00}$ for \kst and $\phi$ vector mesons and its comparison with  theoretical calculation based on $\phi$-meson field~\cite{STAR_spinalign}.}
\label{vector_meson_spin_align}
\end{figure}
The ALICE and STAR collaborations recently measured the $\rho_{00}$ for \kst and $\phi$ in Pb–Pb and Au–Au collisions~\cite{ALICE_spinalign, STAR_spinalign}.  Fig.~\ref{vector_meson_spin_align} shows that the $\rho_{00} <$ 1/3 for \kst and $\phi$ at low $p_{T}$ ($<$ 2~GeV) in semi-central Pb-Pb collisions, whereas the high $p_{T}$ measurements are consistent with $\rho_{00}$ $=$ 1/3~\cite{BMohanty_spinalign}. The deviation of $\rho_{00}$ from 1/3 is highest in semi-central collisions whereas consistent with 1/3 in most central and peripheral collisions, following the impact parameter dependence of the global angular momentum.~\cite{ALICE_spinalign, STAR_spinalign} The $\rho_{00}$ for \kst follows the similar pattern in Au–Au collisions at RHIC~\cite{STAR_spinalign}. These results are consistent with the quark recombination model that attributes the low $p_{T}$ vector meson spin alignment to the quark polarization through spin-orbit coupling and hadronization via recombination of polarized quarks~\cite{spinalign_th_1, BMohanty_spinalign}. The $\rho_{00}$ is consistent with 1/3 for \kzs (spin 0) in semi-central Pb--Pb (Fig.~\ref{vector_meson_spin_align}) and Au--Au collisions, and also for vector mesons in pp collisions where the initial angular momentum is not expected~\cite{ALICE_spinalign, STAR_spinalign}.\\
Interestingly, $\rho_{00}$ $>$ 1/3 for $\phi$ at intermediate $p_{T}$ in Au–Au collisions at RHIC~\cite{STAR_spinalign} as shown in Fig.~\ref{vector_meson_spin_align}. This is inconsistent with the $\rho_{00} (\phi) < 1/3$ at the LHC and quark recombination model estimations~\cite{BMohanty_spinalign}. Recent theory calculation indicates that the coherence $\phi$ meson field can generate $\rho_{00}$ $>$ 1/3 for $\phi$ meson at RHIC~\cite{meson_field_calculation}. The finite $\rho_{00}$ values indicating vector meson spin alignment at RHIC and the LHC are surprisingly large in the context of $\Lambda$ polarization. In a thermal and non-relativistic approach, the hyperon polarization and vector meson spin alignment are associated with the thermal vorticity ($\omega$/T) in the following way: $P_{\Lambda}$ $\simeq$ $\frac{1}{4}\frac{\omega}{T}$ and $\rho_{00}$ $\simeq$ $\frac{1}{3}\left(1-\frac{\left( \omega/T\right)^{2}}{3}\right)$~\cite{BMohanty_spinalign}. At LHC, $P_{\Lambda} \approx$ 0 implies $\rho_{00} \approx$ 1/3 (Fig.~\ref{gpol_lam}). However, the $\rho_{00}$ values in Fig.~\ref{vector_meson_spin_align} are surprisingly large, and local polarization of quarks and anti-quarks may contribute to such large vector meson spin alignment~\cite{BMohanty_spinalign}. More theoretical input is needed for a better understanding of the vector meson spin alignment results at relativistic heavy-ion collisions.

In addition to the global vorticity due to orbital angular momentum, anisotropic expansion of the medium generates non-trivial vorticity fields along different directions~\cite{ALICE_localpol}. For example, in non-central nucleus-nucleus collisions, the strong elliptic flow generates a non-zero vorticity component along the
beam axis ($z$). The vorticity direction and the resulting hyperon polarization exhibit a quadrupole structure in
the transverse plane, hence termed local polarization~\cite{ALICE_localpol}. The $\Lambda$ ($\overline{\Lambda}$) polarization due to elliptic flow-induced vorticity is measured relative to the second harmonic symmetry plane $\psitwo$, and is evaluated as~\cite{ALICE_localpol, STAR_localpol}:
\begin{equation}
\pzstwo = \frac {\mean{{P_{\rm z} \sin(2\varphi - 2 \psitwo)}}} {Res(\psitwo)},
P_{\rm z}
= \frac{\mean{\cos\theta_{\rm p}^{*}}}
{\alpha_{\rm H}\mean{(\cos\theta_{\rm p}^{*})^{2}}};
\label{Eq11}   
\end{equation}
where $\theta_{\rm p}^{*}$ is the polar angle of the
daughter momentum direction in the hyperon rest frame. $\varphi$ is the hyperon azimuthal angle, $\psitwo$ is the reconstructed second harmonic event plane angle, and Res$(\psitwo)$ is the event plane resolution correction~\cite{evplane_res_voloshin}. The factor $\mean{(\cos\theta_{\rm p}^{*})^{2}}$ corrects for finite acceptance along the longitudinal direction. The sign of $\pzstwo$ determines the phase of the $P_{\rm z}$ modulation relative to the second harmonic symmetry plane.\\
\begin{figure}[htb]
\includegraphics[width=4.12cm, height=3.8cm]{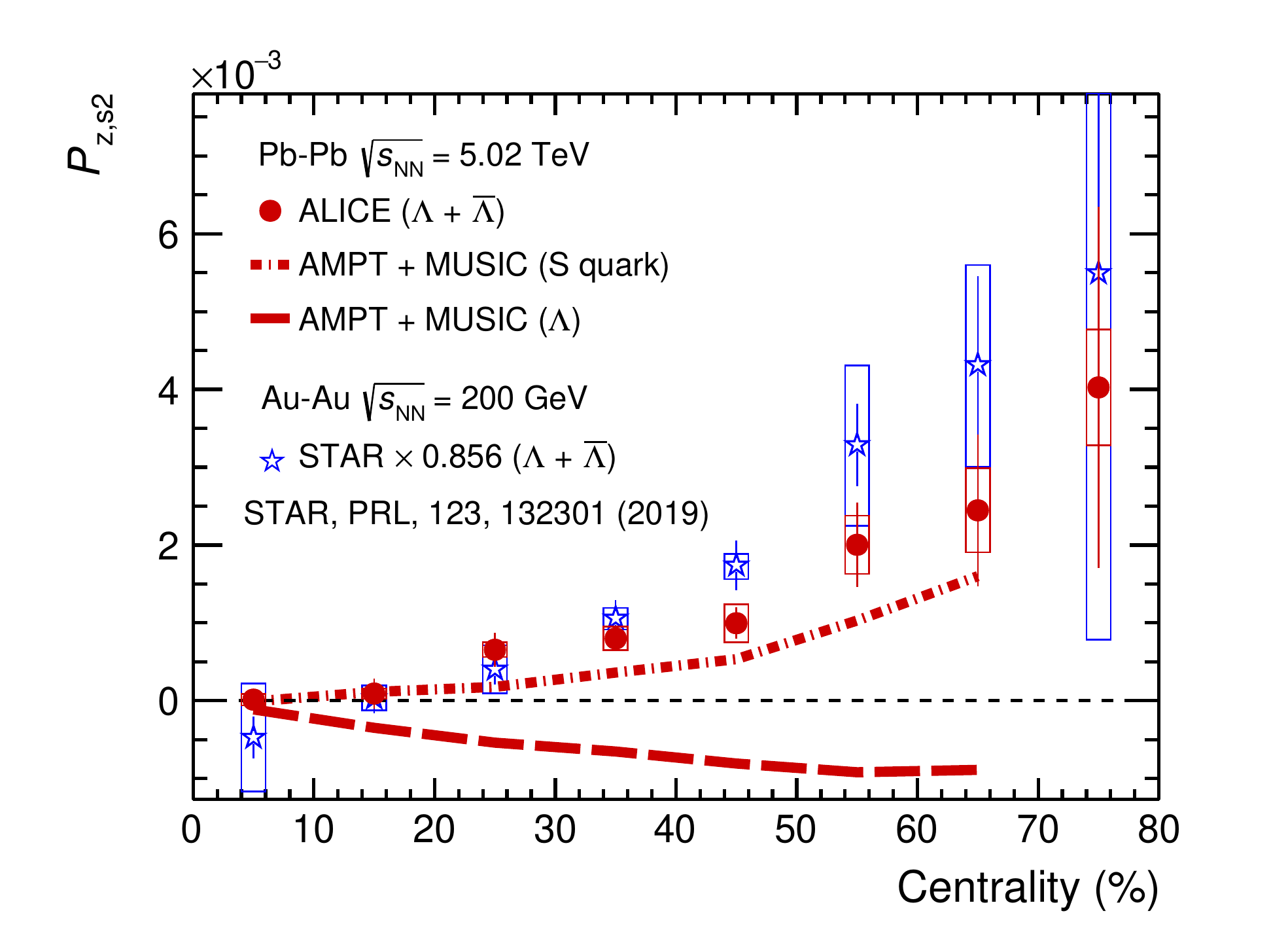}
\includegraphics[width=4.12cm, height=3.8cm]{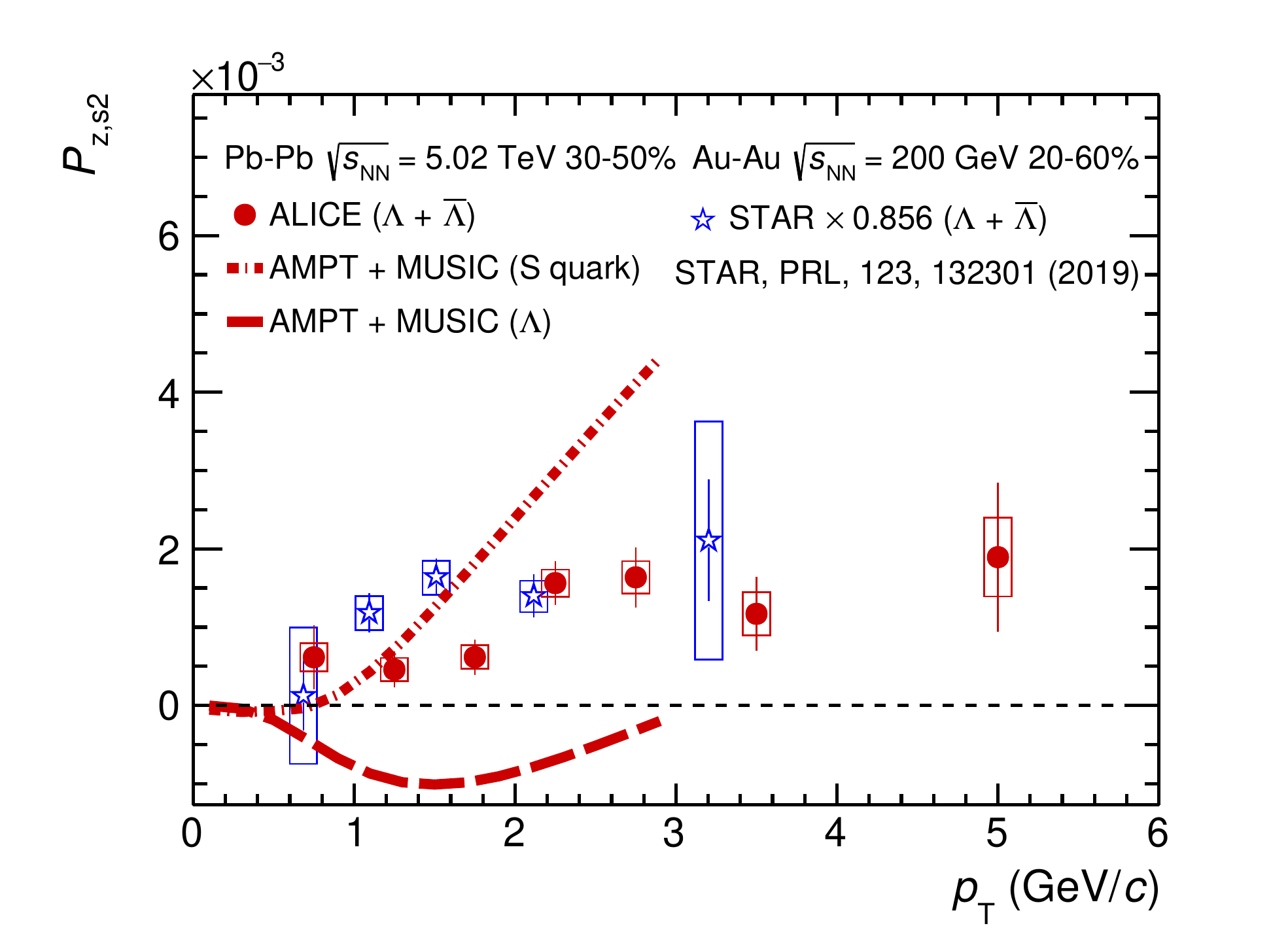}
\includegraphics[width=4.12cm, height=3.8cm]{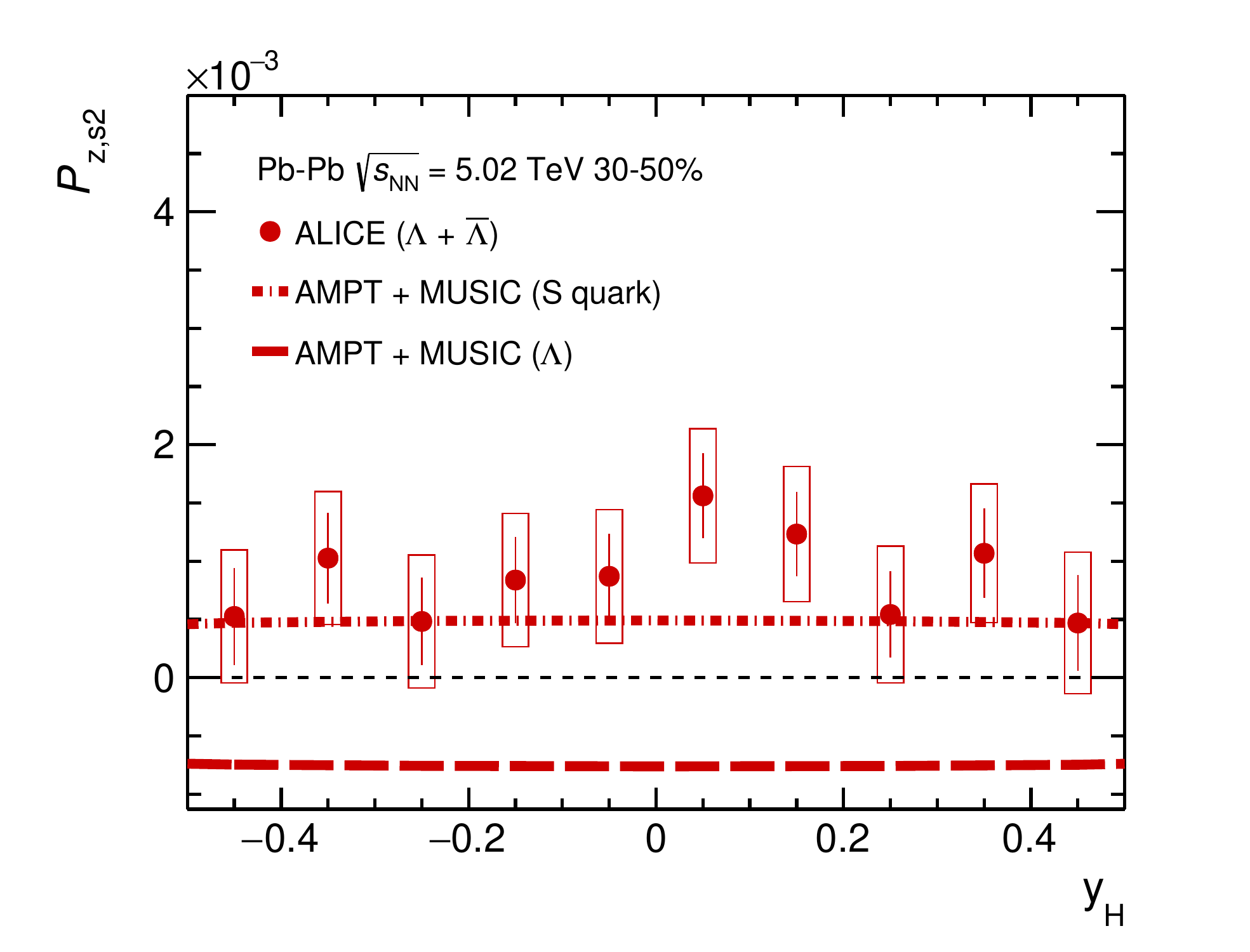}
\caption{ (color online) {\bf Left}: Centrality dependence of $\pzstwo$ and {\bf Middle}: $p_{T}$ dependences of $\pzstwo$ in Pb--Pb collisions at $\sqrt{s_{\rm NN}} =$ 5.02 TeV and Au--Au collisions at $\sqrt{s_{\rm NN}} =$ 200 GeV. {\bf Right}: Rapidity dependence of $\pzstwo$. The AMPT+MUSIC estimations for Pb--Pb collisions at $\sqrt{s_{\rm NN}} =$ 5.02 TeV are shown in all three plots~\cite{ALICE_localpol}.}
\label{lpol_star_alice}
\end{figure}

The centrality and $p_{\rm T}$ dependences of $\pzstwo$ measured by the ALICE Collaboration in Pb--Pb collisions at $\sqrt{s_{_{\rm NN}}} = 5.02$~TeV~\cite{ALICE_localpol} and its comparison with the STAR measurement for Au--Au collisions at $\sqrt{s_{_{\rm NN}}}= 200$~GeV~\cite{STAR_localpol} are shown in Fig.~\ref{lpol_star_alice}. The STAR results are rescaled with a factor of 0.856 to take into account the different $\alpha_{\rm H}$ values used in these measurements~\cite{ALICE_localpol}. The $\pzstwo$ for $\Lambda$  and $\overline\Lambda$ are consistent with each other as expected from the elliptic flow-induced vorticity and combined to calculate the final result. The $\pzstwo$ measured at both $\sqrt{s_{_{\rm NN}}}$ has similar magnitude and phase dependence, and it does not
exhibit a significant dependence on rapidity~\cite{ALICE_localpol}. Fig.~\ref{lpol_star_alice} also shows the comparison between the ALICE results and the $\pzstwo$ values estimated from the fluid shear and thermal vorticity in a 3+1 D hydrodynamical model
(MUSIC) with AMPT initial conditions~\cite{ALICE_localpol, Baochi_shear_lpol}. The model generates positive $\pzstwo$ and qualitatively explains the data assuming constituent strange quark mass as the spin carrier mass. However, the model generates negative $\pzstwo$ in case the hyperon mass is used as the spin carrier mass in the calculation. It is important to note that the thermal vorticity-based hydro and transport calculations explain the $\sqrt{s_{_{\rm NN}}}$ dependence of global polarization ($P_{H}$) reasonably well as shown in Fig.~\ref{gpol_lam}, however, fail to explain the azimuthal angle dependence $P_{\rm H}$~\cite{Baochi_shear_lpol, Becattini_shear_lpol}. The introduction of shear-induced polarization along with additional assumptions on the hadronization temperature or mass of the spin carrier reproduce the experimentally observed positive $\pzstwo$, and the phase of $P_{\rm H}$ at RHIC and the LHC energies~\cite{Baochi_shear_lpol, Becattini_shear_lpol}. A detailed theoretical understanding of the quark spin polarization in the QGP, spin transfer at the hadronization, and the effect of hadronic scattering on the spin polarization are required for a better understanding of the experimental results. The ongoing and future precise measurements of spin polarization and spin alignment in different collision systems and energies would provide further insights into the dynamics of vorticity and particle polarization in heavy-ion collisions.




\end{document}